\documentclass{eas}
\usepackage{graphicx}
\newcommand{\HI}{H\,I\,\,}
\newcommand{\cmm}{cm$^{-2}\:$}
\newcommand{\cmmm}{cm$^{-3}\:$}
\newcommand{\kms}{km~s$^{-1}\:$}
\newcommand{\sun}{\odot}

\begin{document}

\title{A High-Resolution Study of Two Galactic \HI Halo Clouds in the Ophiuchus Superbubble}
\runningtitle{Two \HI Clouds in the Ophiuchus Superbubble}
\author{Y. Pidopryhora}\address{Joint Institute for VLBI in Europe, Postbus 2, 7990 AA, Dwingeloo, The Netherlands}
\author{F. J. Lockman}\address{National Radio Astronomy Observatory, P. O. Box 2, Green Bank, WV 24944, USA}
\author{M. P. Rupen}\address{National Radio Astronomy Observatory, P. O. Box 0, Socorro, NM 87801, USA}
\begin{abstract}
Two compact \HI clouds which seem to belong to the Ophiuchus superbubble were studied at $\sim 30^{\prime\prime}$ resolution using the Very Large Array (VLA) in C and D configurations together with the Green Bank Telescope (GBT) providing the short-spacing flux. Here we present preliminary results of the data analysis.
\end{abstract}
\maketitle
\section{Introduction}
The extra-planar \HI of the Milky Way has been discovered to contain numerous cloud-like structures when observed in the 21~cm line (Simonson \cite{Simonson71}; Lockman \cite{Lockman02}; Lockman \& Pidopryhora \cite{LP}; Ford {\em et al.\/} \cite{Ford08}). These halo clouds have motions consistent with Galactic rotation and do not seem to be related to the classic intermediate- or high-velocity \HI clouds. They are found to distances $>1$~kpc from the plane and at the same time there is evidence (Stil {\em et al.\/} \cite{Stil06}) that their population extends right down into the plane, making them a truly ubiquitous Galactic phenomenon.

While most easily observed by single-dish telescopes (Lockman \cite{Lockman02}; Ford {\em et al.\/} \cite{Ford08}), chosen dense and compact clouds of this kind can also be detected by interferometers such as the VLA with reasonable exposure times of a few hours (Pidopryhora {\em et al.\/} \cite{AAS04}). Such high-resolution studies reveal not only the internal structure of the clouds but also allow us to use them as probes to measure the basic properties of the interstellar medium (ISM) (Lockman \& Pidopryhora \cite{LP}).

\section{Observations and Data Reductions}

A sample of about 20 \HI halo clouds was observed in 21~cm  emission with the VLA in D configuration (Pidopryhora {\em et al.\/} \cite{AAS04}; Lockman \& Pidopryhora \cite{LP}; Pidopryhora \cite{PhD}). Two of these clouds which seem to be parts of the Ophiuchus superbubble (Pidopryhora {\em et al.\/} \cite{PLS}) were also later observed in C configuration. Their Galactic coordinates are $\ell=19^\circ .4$ and $b=+6^\circ .3$ (G19.4+6.3); $\ell=27^\circ .0$, $b=+6^\circ .3$ (G27.0+6.3).

To provide short-spacing flux data (see e.~g. Stanimirovic {\em et al.\/} \cite{Stanimirovic}) additional observations were made with the GBT. After the basic reductions done separately the VLA C and D array data were combined using the DBCON procedure of AIPS and a dirty image of the resulting dataset was made. Then the reductions proceeded in the following two ways: 1.~the dirty image was cleaned with SDI CLEAN method (AIPS's SDCLN) and then combined with the GBT image in Miriad using IMMERGE procedure; 2.~the dirty image was deconvolved using MEM algorithm (AIPS's VTESS) with the GBT image used as the default image. The resulting cubes of both methods were compared and found matching in both the peak and integral flux within a few per cent. In both cases the synthesized beam was $\sim 30^{\prime\prime}$ and rms noise in one channel $\approx 1.5$~mJy/beam $\approx 1.1$~K. The final \HI column density maps are shown in Fig.~1 and a few sample spectra in Fig.~2.

\section{Cloud Properties}

These clouds are very close to the tangent points and thus their distances can be estimated (Pidopryhora {\em et al.\/} \cite{PLS}). This in turn allows us to derive other cloud properties, shown in the Table~1. The linear resolution of the study is
estimated to be $\sim 1$~pc. The densest cloud regions are not resolved and thus the lower limit of their number density can be deduced in assumption of spherical symmetry. Together with the spectral measurements from Fig.~2 it allows us to make an estimate of the core pressure and it too can be considered a lower limit since the linewidth is probably close to its true thermal value.

\section{Discussion}

Both recent experimental (Pidopryhora {\em et al.\/} \cite{PLS}; Dawson {\em et al.\/} \cite{Dawson08}; Ford {\em et al.\/} \cite{Ford08}) and theoretical (Audit \& Hennebelle \cite{Audit05}; Breitschwerdt \& de Avillez \cite{BdA06}; V\'{a}zquez-Semadeni {\em et al.\/} \cite{Enrique07}) studies show that energetic processes in ISM, superbubbles in particular lead to formation of dense and compact quasi-stable cloudlets. Hence it is natural to hypothesize that {\em most} of the corotating Galactic \HI halo clouds are formed in this way. Since it is currently believed (Oey \cite{Oey04}) that superbubbles are a very common phenomenon in Galactic disk-halo transition region, this would explain the persistence of these clouds in the Milky Way's halo. On the other hand formation of the clouds high in the halo eliminates the need to puzzle over their confinement mechanisms (Lockman \& Pidopryhora \cite{LP}).

The look of these two clouds as well as of the other halo clouds studied at high resolution (Pidopryhora \cite{PhD}) seems to indicate that the halo clouds are unstable objects. High core densities and sharp density contrasts probably indicate shocks. There is also a striking left-to-right morphological asymmetry of both G19.4+6.3 and G27.0+6.3, with more diffuse matter to the left, ``inside the bubble'' and sharper density contrast to the right, a possible indication of the outward shock. Thus the morphology of both clouds is consistent with them being pieces of the Ophiuchus superbubble's disintegrating envelope, G19.4+6.3 even seems to display a characteristic Rayleigh-Taylor ``spur''.

We thank Joanne Dawson, Alyson Ford and Enrique V\'{a}zquez-Semadeni for fruitful discussions during the conference.


\begin{center}
{\bf Table 1.} Properties of the Clouds.
  \begin{tabular}{l | c | c | c }
    \hline
    Property & Unit & G19.4+6.3 & G27.0+6.3  \\
    \hline
    distance & kpc & 8.0 & 7.6 \\
    height above the Galactic plane & pc & 890 & 840 \\
    total cloud size & pc & 70 & 70 \\
    total cloud mass (\HI) & $M_{\sun}$ & 1500 & 1300 \\
    peak column density & $10^{20}$~\cmm & 2.6 & 2.8 \\
    core line FWHM & \kms & 3.7 & 2.8 \\
    equivalent kinetic temperature & K & 300 & 170 \\
    estimated core number density & \cmmm & $\geq 75$ & $\geq 87$ \\
    estimated core pressure, $P/k = nT$ & $10^3$~K~\cmmm & 22.5 & 14.8\\
        \hline
  \end{tabular}
\end{center}

\begin{figure}
\includegraphics[width=12.5cm]{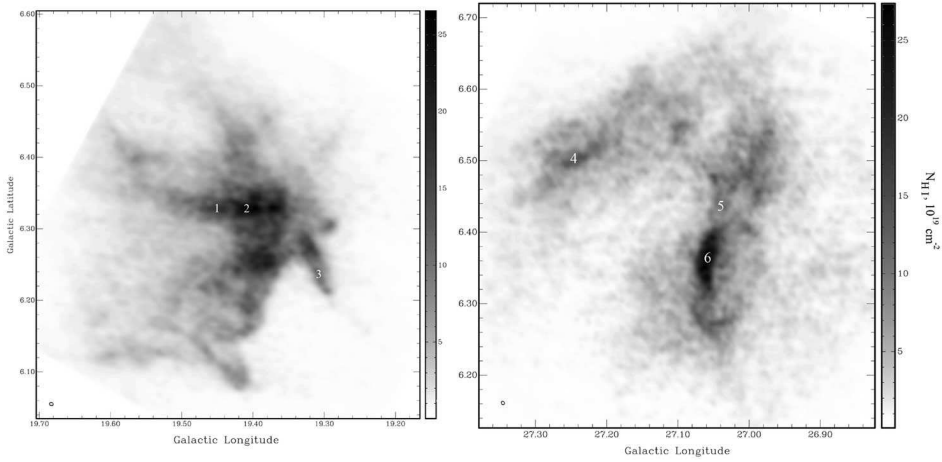}
\caption{Total \HI column density maps of the clouds G19.4+6.3 (left panel, integrated over 48 spectral channels of 0.64~\kms width, rms noise is $8.7 \cdot 10^{18}$~\cmm, synthesized beam FWHM $32.5^{\prime\prime} \times 25.6^{\prime\prime}$, which at the estimated distance of 8.0~kpc is equivalent to the linear resolution of 1.12~pc) and G27.0+6.3 (right panel, integrated over 37 spectral channels of 0.64~\kms width, rms noise is $9.3 \cdot 10^{18}$~\cmm, synthesized beam FWHM $29.7^{\prime\prime} \times 24.5^{\prime\prime}$, which at the estimated distance of 7.6~kpc is equivalent to the linear resolution of 1.05~pc). The panel sizes were chosen to match the angular scale. These maps were produced using the resulting VTESS model of the signal. The synthesized beam is shown in the bottom-left corner of each panel. White numbers mark where the sample spectra shown in Fig.~2 were measured.}
\end{figure}

\begin{figure}
\includegraphics[width=11.68cm]{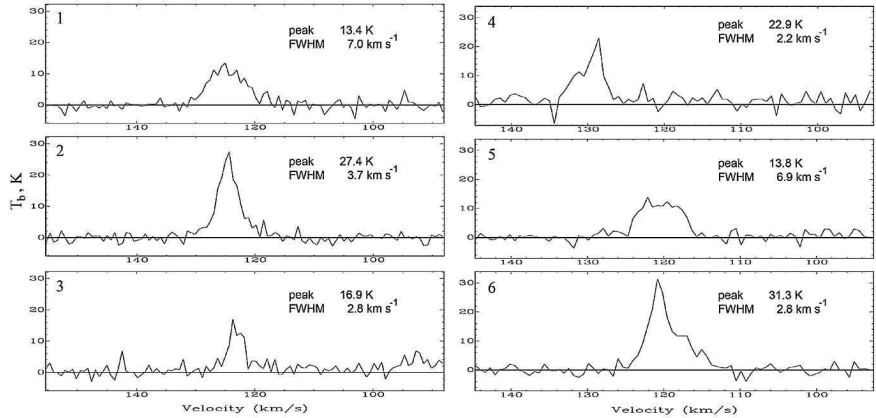}
\caption{Sample spectra of the clouds measured at qualitatively different regions. The places where the spectra were measured are marked by the corresponding numbers in Fig~1. The peak value and FWHM of each spectrum are shown at the top-right corners of the panels.}
\end{figure}

\end{document}